\documentclass [twocolumn,aps,amsmath,amssymb]{revtex4-1}

\usepackage{graphicx}
\usepackage{dcolumn}
\usepackage{bm}
\usepackage{color}

\begin{document}


\title{Localization of 5D Elko Spinors on Minkowski Branes}


\author{Yu-Xiao Liu}
\email{liuyx@lzu.edu.cn}
\affiliation{Institute of Theoretical Physics, Lanzhou University, Lanzhou 730000, People's Republic of China}

\author{Xiang-Nan Zhou}
\email{zhouxn10@lzu.edu.cn, corresponding author}
\affiliation{Institute of Theoretical Physics, Lanzhou University, Lanzhou 730000, People's Republic of China}

\author{Ke Yang}
\email{yangke09@lzu.edu.cn}
\affiliation{Institute of Theoretical Physics, Lanzhou University, Lanzhou 730000, People's Republic of China}

\author{Feng-Wei Chen}
\email{chenfw10@lzu.edu.cn}
\affiliation{Institute of Theoretical Physics, Lanzhou University, Lanzhou 730000, People's Republic of China}



\begin{abstract}
{Recently, a new spin-1/2 fermionic quantum field with mass dimension one in four dimensions--Elko field $\lambda$ was introduced as a candidate of dark matter. In this paper, we investigate the localization of 5D Elko {spinors} on Minkowski branes by presenting the equation of {the} Elko KK modes. For the 5D free massless Elko field, the zero mode can be localized on Randall-Sundrum thin brane but can not be localized on the majority of thick branes. There {do} not exist bound massive KK modes on all these branes. If the 5D mass term is introduced, there will exist bound Elko zero mode in Randall-Sundrum brane model. And when we introduce the Yukawa type coupling $\eta \phi^2 \mathop \lambda\limits^\neg\lambda$ with $\phi$ the background scalar field, the Elko zero mode can be localized on some special thick branes with a particular coupling constant $\eta$. Nevertheless, the massive KK modes still can not be localized on these branes. These results are very different from that of the conventional Dirac spinor {field} and the scalar field.}
\end{abstract}

\maketitle  


\section{Introduction}
\label{section1}
The idea that our world is restricted in a 4D hyper-surface (brane) which is embedded in a multi-dimensional space-time (bulk) drew more and more attentions in recent years. Braneworld theory was originated from the string/M theory. In the framework of braneworld scenarios, Standard Model (SM) fields are bound to the brane, while the gravity can propagate in the bulk. The possibility that extra dimensions can be non-compact
\cite{Rubakov:1983ab,Rubakov:1983cd,Akama:1982jy,Randjbar:1986,Randall:1999ee,Randall:1999vf,Lykken:1999nb} or be large \cite{Antoniadis:1990,ArkaniHamed:1998rs,Antoniadis:1998ig} gives us a novel road to solve some long-standing problems in high-energy physics and cosmology, such as the hierarchy problem, i.e., the large difference between the electro-weak scale $M_{\text{EW}}\sim1$Tev and the Plank scale $M_{\text{Pl}}\sim10^{16}$Tev \cite{ArkaniHamed:1998rs,Antoniadis:1998ig}, and the cosmological constant problem
\cite{Rubakov:1983cd,Randjbar:1986,Kehagias:2004fb,Gogberashvili:1999ad,ArkaniHamed:2000eg,Alcaniz:2002qm,Liu:2010am}. The famous Randall-Sundrum (RS) brane model was presented in the end of 90's \cite{Randall:1999ee,Randall:1999vf}. In the RS brane model, extra dimensions may be non-compact, i.e., the size of extra dimensions can be {infinite}. But the thickness of the ideal RS brane is zero. A more realistic brane should have thickness. The thick brane scenarios are usually based on gravity coupled to a scalar field \cite{DeWolfe,Abdyrakhmanov:2005fs,Gremm:1999pj,Afonso:2006gi,Kehagias:2000au,Bazeia:2008zx}.
Thick brane is naturally generated by a background scalar instead of by introducing a delta function artificially \cite{DeWolfe}. At the same time, the scalar can {provide} the ``materials'' which make the thick brane.  More information about thick brane solutions can be found in the review article \cite{Dzhunushaliev:2009va}.

An important and interesting issue in braneworld scenarios is how are various bulk matter fields localized on branes by a natural mechanism. What we do know is that the gravity \cite{Randall:1999ee,Randall:1999vf,HerreraAguilar:2010kt} and massless scalar field \cite{Bajc:1999mh} can be localized on branes of different types. While the spin-1 Abelian vector fields can be localized on some thick branes and 6D RS brane instead of 5D RS brane \cite{Oda:2000zc,Liu:2009dwa,Liu:2008wd,Liu:2008pi}. The localization of spin-1/2 fermion is very interesting. There can exist a single bound state and a continuous spectrum of massive KK modes with scalar-fermion coupling in some cases \cite{Liu:2007ku,Zhang:2007ii,Bazeia:2008ic,Koroteev:2009xd,Flachi:2009uq,Zhao:2009ja,Li:2010dy,Castro:2010uj,Chumbes:2010xg}. On some other thick branes, there can exist discrete KK modes (mass gaps) and continuous spectrum which starts at a positive value \cite{Liu:2008wd,Liu:2008pi,Zhao:2010mk,Liu:2010pj,Kodama:2008xm,Brihaye:2008am}. In Refs.~\cite{Liu:2009dw,Liu:2009uca}, the spectra of 4D fermions on some symmetric and asymmetric thick
branes and anti-de Sitter thick branes are constituted of bound KK modes. {Furthermore} it was found that there exist fermion resonances on some thick branes, and the life-times of the resonances are decided by the structure of the branes, the Yukawa coupling between the fermionic field and the background scalar field, and the coupling constant \cite{Liu:2010pj,Liu:2009dw,Ringeval:2001cq,Koley:2004at,Davies:2007tq,Liu:2009mga,Almeida:2009jc,Liu:2009ve}.

On the other hand, 
in 2005, Ahluwalia and Grumiller introduced a new quantum field which is a spin-1/2 fermionic quantum field with mass dimension one \cite{Ahluwalia:2004sz,Ahluwalia:2004ab}. It was named as Eigenspinoren des Ladungskonjugationsoperators (Elko) in German, i.e., eigenspinors of the charge conjugation operator. Elko belongs to non-standard Wigner classes \cite{Ahluwalia:2004sz,Gillard:2009zw}
and it will be better to understand Elko in the scope of Very Special Relativity framework \cite{Ahluwalia:2010zn}. One of the consequences of mass dimension one is that Elko can interact with itself, gravity and Higgs doublet, but the mismatch of mass dimensions with Dirac fermions prevents it from entering the fermionic doublets of the SM \cite{Ahluwalia:2004ab}. In addition, Elko is a non-local field and the Lorentz symmetry is broken because there exists a preferred direction. Elko is localized along this direction \cite{Ahluwalia:2008xi,Ahluwalia:2009rh}. Ahluwalia and Grumiller suggested that Elko can be considered as a first-principle candidate of dark matter \cite{Ahluwalia:2004sz,Ahluwalia:2004ab}. Elko also can be used to investigate some cosmological problems such as the horizon problem, the dark energy problem and so on. All of these interesting properties of Elko have attracted more and more attentions \cite{Ahluwalia:2009ia,Gredat:2008qf,Shankaranarayanan:2009sz,Shankaranarayanan:2010st,Wei:2010ad,Wei:2011yr,Boehmer:2009aw,Boehmer:2010ma,Boehmer:2008rz,Chee:2010ju,Rocha:2005,Rocha:2007,Rocha:2008,Rocha:2009aa,Rocha:2009bb,daRocha:2011yr,Fabbri:2009ka,Fabbri:2009aj,Fabbri:2010va,Fabbri:2010ws,Fabbri:2010qv,Fabbri:2011mi,Fabbri:2012yg,Dias:2010aa,Boehmer:2006qq,Lee:2010nf,Basak:2011wp}.
Therefore, Elko is a new matter field which we cannot ignore.

In the framework of braneworld scenarios,
the localization of various matter fields except Elko on branes has been studied and the mass spectra also have been given.
All of these researches about Elko motivate us to investigate the interesting problem
that whether higher dimensional Elko field can be localized on various kinds of branes. The peculiar properties of Elko may result in that its localization is very different from the ones of other matter fields.
At the same time, among these brane models, the Minkowski (flat) one is the simplest brane model. For the first investigation about the localization of the new matter field, we choose the Minkowski branes as our subjects. We will show in this paper that, only the Elko zero mode, i,e, the 4D massless Elko field can be localized on RS {brane} and on some special thick branes with coupling term. There will not exist bound massive KK modes on these Minkowski branes. The conclusion is very different from the ones of other SM matter fields about localization on branes. It tells us that the coupling between {4D} Elko and Higgs which can generate the mass of 4D Elko \cite{Ahluwalia:2004ab} is crucial.

The organization of the paper is as follows: In Sec.~\ref{section2}, we first briefly review the Elko quantum field. Then, in Sec.~\ref{section3}, we discuss the localization of {a} 5D free massless Elko field on various Minkowski branes by presenting the equation of the Elko KK modes. And in Sec.~\ref{section4}, we discuss the localization of {a} 5D Elko with coupling term on these Minkowski branes. Then we will list the advantages of choosing Elko as the candidate of dark matter in Sec.~\ref{section5}. These advantages give us the motivation of investigation of Elko's localization on branes. Finally, the conclusion is given.

\section{Review of Elko field}
\label{section2}

Elko can not be expressed in Weinberg's formalism and it belongs to non-standard Wigner classes
\cite{Ahluwalia:2004sz,Gillard:2009zw}. Elko can originate from Very Special Relativity \cite{Ahluwalia:2010zn} and obey the unusual property $(CPT)^2=-\mathbb{I}$.  Here charge conjugation $C$ {is} defined as
\begin{eqnarray}
C=\left( \begin{array}{cc}
\mathbb{O}   & i\Theta \\
-i\Theta     & \mathbb{O}
\end{array} \right)K,
\end{eqnarray}
where $K$ is the complex conjugation operator, and $\Theta$ is the spin one half Wigner time reversal operator satisfying $\Theta(\vec{\sigma}/2)\Theta^{-1}=-(\vec{\sigma}/2)^{*}$. Thus, the $\Theta$ {is} given by
\begin{eqnarray}
\Theta=\left( \begin{array}{cc}
0 & -1 \\
1 & 0
\end{array} \right).
\end{eqnarray}

Elko spinors are eigenspinors of the charge conjugation operator: $C\lambda(k^\mu)=\pm\lambda(k^\mu)$ ($k^\mu$ is a polarization vector). The plus sign generates the self-conjugate spinors which are denoted by $\varsigma(k^\mu)$ and the minus sign generates the anti-self-conjugate spinors which are denoted by $\tau(k^\mu)$. In addition, we indicate the two possible helicity eigenstates with $\chi_{\pm}(k^\mu)$, then the four types of Elko can be written as
\begin{eqnarray}
\varsigma_\pm(k^\mu) &=& ~~{\text{i}\Theta[\chi_{\pm}(k^{\mu})]^{*} \choose \chi_\pm(k^{\mu})}, \\
\tau_\pm(k^\mu) &=&\pm{-\text{i}\Theta[\chi_{\mp}(k^{\mu})]^{*} \choose \chi_\mp(k^{\mu})}.~~~~
\end{eqnarray}
Here $\chi_{\pm}(k^{\mu})$ can be read as
\begin{eqnarray}
\chi_{+}(k^{\mu})=e^{-\text{i}\phi/2}\sqrt{m}{1 \choose 0},~~\chi_{-}(k^{\mu})=e^{\text{i}\phi/2}\sqrt{m}{0 \choose 1}.~~\label{chim}
\end{eqnarray}
{
It can always transform $k^{\mu}$ as $p^{\mu}$ by a transformation operator $\Gamma$ ($p^{\mu}$ is a general vector and represents ($E$, $p_x$, $p_y$, $p_z$)), and $\lambda(p^{\mu})$ is also an Elko. The $\Gamma$ is given by \cite{Ahluwalia:2010zn}
\begin{eqnarray}
\Gamma=\left( \begin{array}{cccc}
\sqrt{\frac{m}{E-p_z}} & \frac{p_x-\text{i}p_y}{\sqrt{m(E-p_z)}} & 0                                         & 0 \\
                 0     & \sqrt{\frac{E-p_z}{m}}                  & 0                                         & 0 \\
                 0     &                  0                      & \sqrt{\frac{E-p_z}{m}}                    & 0 \\
                 0     &                  0                      & -\frac{p_x+\text{i}p_y}{\sqrt{m(E-p_z)}}  & \sqrt{\frac{m}{E-p_z}}
\end{array} \right).~~~
\end{eqnarray}
One can get 4D massless Elko from the form of $\lambda(p^{\mu})$ by taking the massless limit, and the $\varsigma_{-}(p^\mu)$ and $\tau_{+}(p^\mu)$ will vanish but $\varsigma_{+}(p^\mu)$ and $\tau_{-}(p^\mu)$ will not in the massless limit.}

The dual spinors for Elko are defined as
\begin{eqnarray}
  {\mathop {\varsigma}\limits^\neg}_{\pm}(p^{\mu})=\pm[\varsigma_{\mp}(p^{\mu})]^{\dag}\gamma^{0},~~
  {\mathop {\tau}\limits^\neg}_{\pm}(p^{\mu})=\pm[\tau_{\mp}(p^{\mu})]^{\dag}\gamma^{0}.
\end{eqnarray}
When the Dirac operator $\gamma_{\mu}p^{\mu}$ acts on Elko spinors, the results are
\begin{eqnarray}
  \gamma_{\mu}p^{\mu}\varsigma_{\pm}(p^{\mu})=\mp m\varsigma_{\mp}(p^{\mu}),~
  \gamma_{\mu}p^{\mu}\tau_{\pm}(p^{\mu})=\pm m\tau_{\mp}(p^{\mu}).~~~~
\end{eqnarray}
Here we choose the gamma matrices $\gamma^\mu$ and $\gamma^5$ in the following form:
\begin{eqnarray}
\gamma^0=\left( \begin{array}{cc}
     \mathbb{O}    & -\text{i}\mathbb{I} \\
    -\text{i}\mathbb{I}    & \mathbb{O}
\end{array} \right),~
\gamma^i=\left( \begin{array}{cc}
    \mathbb{O}    & \text{i}\sigma^i \\
    -\text{i}\sigma^i      & \mathbb{O}
\end{array} \right),~
\gamma^5=\left( \begin{array}{cc}
    \mathbb{I}    & \mathbb{O} \\
    \mathbb{O}    & -\mathbb{I}
\end{array} \right).~~~~\label{representation of gamma matrices}
\end{eqnarray}
It is clear that $\gamma^{\mu}$ satisfy the relation: $\{\gamma^\mu,\gamma^\nu\}=2\eta^{\mu\nu}\mathbb{I}$ with $\eta^{\mu\nu}=\text{diag}(-,+,+,+)$. And when $\gamma^5$ acts on the four types of Elko one gets
\begin{eqnarray}
\gamma^5\varsigma_{\pm}(p^{\mu})=\pm\tau_{\mp}(p^\mu),\qquad \gamma^5\tau_{\pm}(p^\mu)=\mp\varsigma_\mp(p^{\mu}).
\end{eqnarray}
Making use of the Fourier transformation we can obtain the following equations:
\begin{eqnarray}
\gamma^\mu\partial_\mu\varsigma_\pm(x)&=&\mp \text{i}m\varsigma_\mp(x),~\gamma^\mu\partial_\mu\tau_\pm(x)=\pm \text{i}m\tau_\mp(x); ~~~~ \label{5D Dirac operator on Elko spinors}\\
\gamma^5\varsigma_{\pm}(x)&=&\pm\tau_{\mp}(x),\qquad~~\gamma^5\tau_{\pm}(x)=\mp\varsigma_\mp(x). \label{5D gamma 5 on Elko spinors}
\end{eqnarray}
As we know, it should be $\gamma^{\mu}\partial_\mu\psi\propto\psi$ if $\psi$ is a Dirac spinor. Therefore, Elko spinors do not satisfy the Dirac equation. On the other hand, via further discussion, it is found that Elko satisfies the Klein-Gordon (KG) equation: $(\eta^{\mu\nu}\partial_\mu\partial_\nu-m^{2})\lambda(x)=0$. Thus, we get the Lagrangian density of a free Elko in 4D flat space-time: $\mathfrak{L}_{\text{Elko}}=-\frac{1}{2}\partial^{\mu}\mathop\lambda\limits^\neg\partial_{\mu}\lambda-\frac{1}{2}m^{2}\mathop \lambda\limits^\neg\lambda$. For a general curved space-time, the Lagrangian density should be written as \cite{Ahluwalia:2004ab,Gredat:2008qf,Wei:2010ad}
\begin{eqnarray}
  \mathfrak{L}_{\text{Elko}}=-\frac{1}{2}\left[\frac{1}{2}g^{\mu\nu}(\mathfrak{D}_{\mu}
  \mathop \lambda\limits^\neg\mathfrak{D}_{\nu}\lambda+\mathfrak{D}_{\nu}
  \mathop \lambda\limits^\neg\mathfrak{D}_{\mu}\lambda)\right]-
  V(\mathop \lambda\limits^\neg\lambda),~~~~~~
  \label{Elko's Lagrangian density}
\end{eqnarray}
where $V(\mathop \lambda\limits^\neg\lambda)$ is the potential of the Elko field, and $\mathfrak{D}_{\mu}$ represents covariant derivative.

\section{The localization of 5D free massless Elko spinors}
\label{section3}

In this section, we will study the localization of free massless Elko spinors on Minkowski branes in 5D space-time. The metric describing a 4D Minkowski brane embedded in a 5D bulk is generally assumed as
\begin{eqnarray}
ds^{2}=\text{e}^{2A(y)}\eta_{\mu\nu}dx^{\mu}dx^{\nu}+dy^{2}, \label{the line-element for 5D space time}
\end{eqnarray}
where $\text{e}^{2A(y)}$ is the warp factor and $y$ the extra coordinate. Further, by performing the coordinate transformation
\begin{eqnarray}
dz=\text{e}^{-A(y)}dy, \label{coordinate transformation}
\end{eqnarray}
the metric (\ref{the line-element for 5D space time}) transforms to a conformally flat one
\begin{eqnarray}
ds^{2}=\text{e}^{2A}(\eta_{\mu\nu}dx^{\mu}dx^{\nu}+dz^{2}), \label{conformally flat line-element}
\end{eqnarray}
which is more convenient for discussing the localization of gravity and various matter fields.

The action of a free massless Elko field $\lambda$ in 5D space-time should be
\begin{eqnarray}
S_{\text{Elko}}=\int d^{5}x\sqrt{-g}\mathfrak{L}_{\text{Elko}}. \label{Elko's Lagrangina in 5D}
\end{eqnarray}
Here the Lagrangian density for the Elko field is
\begin{eqnarray}
\mathfrak{L}_{\text{Elko}}=-\frac{1}{2}\left[\frac{1}{2}g^{MN}\big(\mathfrak{D}_{M}\mathop \lambda\limits^\neg\mathfrak{D}_{N}\lambda+\mathfrak{D}_{N}\mathop \lambda\limits^\neg\mathfrak{D}_{M}\lambda\big)\right].~~~~ \label{Lagrangian for Elko in 5D}
\end{eqnarray}
In this paper, $M,N\cdots=0,1,2,3,5$ and $\mu,\nu\cdots=0,1,2,3$ denote the 5D and 4D space-time indices, respectively, and $\bar{A},\bar{B}\cdots=0,1,2,3,5$ and $a,b\cdots=0,1,2,3$ denote the 5D and 4D local Lorentz indices, respectively. The covariant derivatives are
\begin{eqnarray}
\mathfrak{D}_{M}\lambda=(\partial_{M}+\Omega_{M})\lambda,~~\mathfrak{D}_{M}\mathop \lambda\limits^\neg=\partial_{M}\mathop \lambda\limits^\neg-\mathop \lambda\limits^\neg\Omega_{M}, \label{covariant derivatives}
\end{eqnarray}
where the tangent space connection $\Omega_{M}$ is defined as
\begin{eqnarray}
   \Omega_{M}&=&-\frac{i}{2}\left(e_{\bar{A}P}e_{\bar{B}}^{~~N} \Gamma^{P}_{MN}
                             -e_{\bar{B}}^{~~N}\partial_{M}e_{\bar{A}N}\right)
                  S^{\bar{A}\bar{B}}, \\
   S^{\bar{A}\bar{B}}&=&\frac{i}{4}[\gamma^{\bar{A}},\gamma^{\bar{B}}].
    \label{tangent space connection}
\end{eqnarray}
Here $e^{\bar{A}}_{~M}$ is the vierbein and satisfies the orthonormality relation $g_{MN}=e^{\bar{A}}_{~M}e^{\bar{B}}_{~N}\eta_{\bar{A}\bar{B}}$. We define $\gamma^M$ as the 5D flat gamma matrixes which satisfy $\{\gamma^{M},\gamma^{N}\}=2\eta^{MN}\mathbb{I}$ with $\eta^{MN}=\text{diag}(-,+,+,+,+)$. The representation of the 5D flat gamma matrixes $\gamma^M$ is the same as the one of 4D flat gamma matrixes $\gamma^\mu$ which is given by (\ref{representation of gamma matrices}). Form (\ref{conformally flat line-element}) the vierbein is given by
\begin{eqnarray}
 e^{\bar{A}}_{~M}=\left(
 \begin{array}{cc}
    \text{e}^{A} \hat{e}^{a}_{~\mu} & 0 \\
      0                             &   \text{e}^{A}
 \end{array} \right),
 \quad  \hat{e}^{a}_{~\mu} = \mathbb{I}.
 \label{vierbein}
\end{eqnarray}
So the non-vanishing components of the spin connection $\Omega_{M}$ for a flat brane are:
\begin{eqnarray}
\Omega_{\mu}=\frac{1}{2}\partial_{z}A\gamma_{\mu}\gamma_{5}. \label{spin connection for flat brane}
\end{eqnarray}

Since the Lagrangian density for the Elko field is similar to the {one for the} scalar field, the equation of motion for Elko is just like the { one for the} scalar field as expected:
\begin{eqnarray}
\frac{1}{\sqrt{-g}}\mathfrak{D}_{M}(\sqrt{-g}g^{MN}\mathfrak{D}_{N}\lambda)=0. \label{Elko's motion equation}
\end{eqnarray}
By considering the conformally flat metric (\ref{conformally flat line-element}) and using the non-vanishing components of the spin connection (\ref{spin connection for flat brane}), we can rewrite Eq. (\ref{Elko's motion equation})
as:
\begin{eqnarray}
  \frac{1}{\sqrt{-g}}\hat{\mathfrak{D}}_{\mu}(\sqrt{-g}\hat{g}^{\mu\nu}\hat{\mathfrak{D}}_{\nu}\lambda)
 +\bigg[-\frac{1}{4}{A'}^{2}\hat{g}^{\mu\nu}\gamma_{\mu}\gamma_{\nu}\lambda\nonumber\\
 +\frac{1}{2}A'\Big(\hat{\mathfrak{D}}_{\mu}(\hat{g}^{\mu\nu}\gamma_{\nu}\gamma_{5}\lambda)
 +\hat{g}^{\mu\nu}\gamma_{\mu}\gamma_{5}\hat{\mathfrak{D}}_{\nu}\lambda\Big)
 \nonumber\\
 +\text{e}^{-3A}\partial_{z}(\text{e}^{3A}\partial_{z}\lambda)\bigg]=0.
  \label{ElkoMotionEq2}
\end{eqnarray}
Here $\hat{g}_{\mu\nu}$ is the induced metric on the brane, and $\hat{\mathfrak{D}}_{\mu}\lambda=(\partial_{\mu}+\hat{\Omega}_{\mu})\lambda$ with $\hat{\Omega}_{\mu}$ the spin connection constructed by the induced metric $\hat{g}_{\mu\nu}$.

For the case of flat branes considered here, $\hat{g}_{\mu\nu}={\eta}_{\mu\nu}$ and hence $\hat{\mathfrak{D}}_{\mu}=\partial_{\mu}$ and the equation of motion {(\ref{ElkoMotionEq2})} can be simplified as
\begin{eqnarray}
 \partial^{\mu}\partial_{\mu}\lambda
  -\! A'\gamma^{5}\gamma^{\mu}\partial_{\mu}\lambda
   \!-\! {A'}^{2}\lambda \!+\!\text{e}^{-3A}\partial_{z}(\text{e}^{3A}\partial_{z}\lambda)=0.~~
  \label{Elko's motion equation for flat brane}
\end{eqnarray}
The existence of the term $-A'\gamma^{5}\gamma^{\mu}\partial_{\mu}\lambda$ in Eq. (\ref{Elko's motion equation for flat brane}) suggests that the general solution would inevitably be a linear combination of two different types of an Elko spinor, which is based on Eqs. (\ref{5D Dirac operator on Elko spinors}) and (\ref{5D gamma 5 on Elko spinors}). Hence, we first decompose the { Elko field} as $\lambda=\lambda_{+} + \lambda_{-}$, and then make the general KK decomposition:
\begin{eqnarray}
\lambda_{\pm}\equiv\text{e}^{-3A/2}\sum_n\left(\alpha_{n}(z)\varsigma^{(n)}_{\pm}(x)+\beta_{n}(z)\tau^{(n)}_{\pm}(x)\right). \label{KK decomposition}
\end{eqnarray}
Here for simplicity we omit the $\pm$ subscript for the $\alpha$ and $\beta$ functions. $\varsigma^{n}_{\pm}(x)$ and $\tau^{(n)}_{\pm}(x)$ are linear independant 4D Elko spinors and satisfy the 4D massive KG equations: $\partial^{\mu}\partial_{\mu}\varsigma^{(n)}_{\pm}=m_{n}^{2}\varsigma^{(n)}_{\pm}$ and $\partial^{\mu}\partial_{\mu}\tau^{(n)}_{\pm}=m_{n}^{2}\tau^{(n)}_{\pm}$.
Note that $\lambda_+$ and $\lambda_-$ are linear independant and the operators in Eq. (\ref{Elko's motion equation for flat brane}) do not change the subscripts ``$+$" and ``$-$".
We find that the equations for ($\alpha_{n+},~\beta_{n+}$) and ($\alpha_{n-},~\beta_{n-}$) are the same. So we just need to consider the ($\alpha_{n+},~\beta_{n+}$) case, which is given by
\begin{eqnarray}
&&\left(\alpha_{n}''-\frac{3}{2}A''\alpha_{n}-\frac{13}{4}(A')^2\alpha_{n}+m_{n}^2\alpha_{n}-\text{i}m_{n}A'\beta_{n}\right)\varsigma^{(n)}_{+}\nonumber\\
&+&\left(\beta_{n}''-\frac{3}{2}A''\beta_{n}-\frac{13}{4}(A')^2\beta_{n}+m_{n}^2\beta_{n}-\text{i}m_{n}A'\alpha_{n}\right)\tau^{(n)}_{+}\nonumber\\
&=&0. \label{KKequation3}
\end{eqnarray}
Here we omit the summation symbol and the coordinate symbols. Then, by linear independance of {the} $\varsigma^{(n)}_{+}$ and $\tau^{(n)}_{+}$, we would arrive at the following equations of motion for the pair $\alpha_{n}$ and $\beta_{n}$:
\begin{eqnarray}
\alpha_{n}''-\left(\frac{3}{2}A''+\frac{13}{4}(A')^{2}-m_{n}^2\right)\alpha_{n}-\text{i}m_{n}A'\beta_{n}&=&0, ~~~~\label{equation of alpha}\\
\beta_{n}''-\left(\frac{3}{2}A''+\frac{13}{4}(A')^{2}-m_{n}^2\right)\beta_{n}-\text{i}m_{n}A'\alpha_{n}&=&0.~~~~~ \label{equation of beta}
\end{eqnarray}
Now we define $a_n(z)$ and $b_{n}(z)$ satisfying the following relations:
\begin{eqnarray}
\alpha_{n}=\frac{1}{\sqrt{2}}(a_n+b_n),\qquad \beta_{n}=\frac{1}{\sqrt{2}}(a_n-b_n).
\end{eqnarray}
It is obvious that $a_n$ and $b_n$ satisfy
\begin{eqnarray}
a_{n}''-\left(\frac{3}{2}A''+\frac{13}{4}(A')^{2}-m_{n}^2+\text{i}m_{n}A'\right)a_{n}=0,\\
b_{n}''-\left(\frac{3}{2}A''+\frac{13}{4}(A')^{2}-m_{n}^2-\text{i}m_{n}A'\right)b_{n}=0.
\end{eqnarray}
For the 5D dual Elko spinor $\mathop \lambda\limits^\neg$, we have the following KK decomposition:
\begin{eqnarray}
{\mathop \lambda\limits^\neg}_{\pm}\equiv\text{e}^{-3A/2}\sum_{n}\left(\alpha_{n}^{*}(z){\mathop \varsigma\limits^\neg}^{(n)}_{\pm}(x)+\beta_{n}^{*}(z){\mathop \tau\limits^\neg}^{(n)}_{\pm}(x)\right). \label{KK decomposition2}
\end{eqnarray}
And the 4D dual Elko spinors satisfy the equations:
\begin{eqnarray}
\partial_{\mu}{\mathop \varsigma\limits^\neg}^{(n)}_{\pm}\gamma^{\mu}&=&\pm\text{i}m_{n}{\mathop \varsigma\limits^\neg}^{(n)}_{\mp},~~ \partial_{\mu}{\mathop \tau\limits^\neg}^{(n)}_{\pm}\gamma^{\mu}=\mp\text{i}m_{n}{\mathop \tau\limits^\neg}^{(n)}_{\mp};\\
{\mathop \varsigma\limits^\neg}^{(n)}_{\pm}\gamma^5&=&\mp{\mathop \tau\limits^\neg}^{(n)}_{\mp},\qquad ~~~~
{\mathop \tau\limits^\neg}^{(n)}_{\pm}\gamma^5=\pm{\mathop \varsigma\limits^\neg}^{(n)}_{\mp}.
\end{eqnarray}
Then by substituting the KK decompositions (\ref{KK decomposition}) and (\ref{KK decomposition2}) into the action (\ref{Elko's Lagrangina in 5D}), and using  Eqs. (\ref{equation of alpha}) and (\ref{equation of beta}), we carry out the KK reduction. For the purpose of getting the action of the 4D massless and massive Elko fields from the action of a 5D free massless Elko:
\begin{eqnarray}
S_{\text{Elko}}&=&-\frac{1}{4}\int d^5x\sqrt{-g}g^{MN}(\mathfrak{D}_M\mathop \lambda\limits^\neg\mathfrak{D}_{N}\lambda+\mathfrak{D}_N\mathop \lambda\limits^\neg\mathfrak{D}_{M}\lambda)\nonumber\\
               &=&-\frac{1}{2}\sum_{n}\int d^4x(\partial^{\mu}\hat{{\mathop \lambda\limits^\neg}}^{n}\partial_{\mu}\hat{\lambda}^{n}+m_{n}^2\hat{{\mathop \lambda\limits^\neg}}^{n}\hat{\lambda}^{n}),
\end{eqnarray}
where $\hat{\lambda}^{n}$ are the 4D general Elko spinors, we should introduce the following orthonormality conditions for $\alpha_{n}$ and $\beta_{n}$:
\begin{eqnarray}
\int \alpha^{*}_{n}\alpha_{m}dz&=&\delta_{nm},\label{orthonormality relation 1}\\
\int \beta_{n}^{*}\beta_{m}dz&=&\delta_{nm},\\
\int \alpha^{*}_{n}\beta_{m}dz&=&\int \alpha_{n}\beta^{*}_{m}dz=\delta_{nm}.
\end{eqnarray}
From these orthonormality relations one can get the {corresponding} relations for $a_{n}$ and $b_{n}$:
\begin{eqnarray}
\int a^{*}_{n}a_{m}dz&=2&\int (\alpha^{*}_{n}+\beta^{*}_{n})(\alpha_{m}+\beta_{m})dz=8\delta_{nm},\\
\int b^{*}_{n}b_{m}dz&=2&\int (\alpha^{*}_{n}-\beta^{*}_{n})(\alpha_{m}-\beta_{m})dz=0. \label{orthonormality relation 2}
\end{eqnarray}
The orthonormality relation of $b_{n}$ indicates that $b_{n}=0$ and $\alpha_{n}=\beta_{n}$. The result is interesting and it means that the KK modes of different types of an Elko spinor are the same and indistinguishable. We can not distinguish  different types of an Elko spinor by just considering their KK modes. Now the KK decomposition of the { 5D Elko field} is
\begin{eqnarray}
\lambda_{\pm}&=&\text{e}^{-3A/2}\sum_{n}\left(\alpha_{n}(z)\varsigma^{(n)}_{\pm}(x)+\alpha_{n}(z)\tau^{(n)}_{\pm}(x)\right)\nonumber\\
             &=&\text{e}^{-3A/2}\sum_{n}\alpha_{n}(z)\hat{\lambda}_{\pm}^{n}(x),\label{Elkodecomposition}
\end{eqnarray}
and the equation of the KK mode $\alpha_{n}$ reads
\begin{eqnarray}
\alpha_{n}''-\left(\frac{3}{2}A''+\frac{13}{4}(A')^{2}-m_{n}^2+\text{i}m_{n}A'\right)\alpha_{n}=0.\label{KKequationforElko}
\end{eqnarray}

\subsection{The localization of the zero mode of a 5D free massless Elko}
\label{secVolcanoPotentials}

We first focus on the localization of the zero mode of a 5D free massless Elko spinor on Minkowski branes.
For the zero mode $\alpha_{0}$, i.e, the 4D massless Elko spinor, Eq. (\ref{KKequationforElko}) is simplified as
\begin{eqnarray}
[-\partial_{z}^{2}+V_{0}(z)]\alpha_{0}(z)=0, \label{KKequationforzeromode}
\end{eqnarray}
where $V_{0}$  is given by
\begin{eqnarray}
V_{0}(z)=\frac{3}{2}A''+\frac{13}{4}{A'}^{2}, \label{effective potential Vz}
\end{eqnarray}
and the orthonormality condition is given by
\begin{eqnarray}
\int \alpha^*_0\alpha_0dz=1.\label{orthonormality relation for zero mode}
\end{eqnarray}
Next we will consider thin and thick Minkowski brane solutions respectively, and analyze the localization of the  Elko zero mode on these branes by using Eqs. (\ref{KKequationforzeromode}) and (\ref{orthonormality relation for zero mode}).

\subsubsection{The thin brane}
\label{zeromodeonthinbrane}

As the typification of thin brane models, we consider the RS model, and investigate the localization of the zero mode of Elko on the RS brane.

In 1999, Randall and Sundrum presented the famous RS model to solve the hierarchy problem \cite{Randall:1999ee}. There are two types of RS model: the RS$\mathrm{I}$ and the RS$\mathrm{II}$. The extra dimension is compact in RS$\mathrm{I}$ model so that the zero mode of Elko is indeed a bound mode in this case. So, we give our attention to the RS$\mathrm{II}$ model with a non-compact extra dimension.

The action in RS$\mathrm{II}$ model is \cite{Randall:1999vf}
\begin{eqnarray}
S&=&S_{\text{gravity}}+S_{\text{brane}},\nonumber\\
S_{\text{gravity}}&=&\int d^{4}x\int dy\sqrt{-G}\left(-\Lambda+\frac{1}{2}R\right),\nonumber\\
S_{\text{brane}}&=&\int d^{4}x\sqrt{-g_{\text{brane}}}\left(V_{\text{brane}}+\mathfrak{L}_{\text{brane}}\right), \label{actionofRS}
\end{eqnarray}
where $R$ is  the 5D Ricci scalar, $G_{MN}$ is the 5D metric, $\Lambda$ and $V_{\text{brane}}$ are cosmological terms in the bulk and boundary, respectively. $G_{MN}$ is given by (\ref{the line-element for 5D space time}). The extra dimension $y$ is non-compact and the solution of the warp factor $A(y)$ is given by
\begin{eqnarray}
A(y)=-k|y|,
\end{eqnarray}
where $k$ is a positive real constant. This solution holds when the boundary and bulk cosmological terms are related by \cite{Randall:1999vf}
\begin{eqnarray}
V_{\text{brane}}=6k,\qquad\Lambda=-6k^2.
\end{eqnarray}
Working with the conformal metric (\ref{conformally flat line-element}), the coordinate transformation (\ref{coordinate transformation}) gives that $k|z|+1=\text{e}^{k|y|}$, and the $V_{0}$ (\ref{effective potential Vz}) has the following form:
\begin{eqnarray}
V_{0}=\frac{19k^2}{4(1+k|z|)^2}-\frac{3k\delta(z)}{1+k|z|}.
\end{eqnarray}
The general solution of {Eq.} (\ref{KKequationforzeromode}) is given by:
\begin{eqnarray}
\alpha_{0}(z)=C_{1}(k|z|+1)^{\frac{1}{2}+\sqrt{5}}+C_{2}(k|z|+1)^{\frac{1}{2}-\sqrt{5}}{,} \label{FreeRS}
\end{eqnarray}
where $C_{1}$, $C_{2}$ are integral parameters. In order to get localized Elko zero mode $\alpha_{0}$ on the thin brane, the orthonormality condition (\ref{orthonormality relation for zero mode}) should be satisfied,  which indicates that $\alpha_{0}(z)$ must be vanished when $z\rightarrow\pm\infty$. The first term of solution (\ref{FreeRS}) will be divergent when $z\rightarrow\pm\infty$, so $C_{1}$ should vanish. $C_{2}$ can be determined according to {the} requirement of the orthonormality condition. Then we get the bound zero mode for {a} 5D massless Elko field on the RS$\mathrm{II}$ brane:
\begin{eqnarray}
\alpha_{0}(z)=\sqrt{(-1+\sqrt{5})k}~(k|z|+1)^{\frac{1}{2}-\sqrt{5}}.
\end{eqnarray}
So the zero mode of {a} 5D free Elko can be localized on RS$\mathrm{II}$ brane.

\subsubsection{The thick brane}
\label{zeromodeonthickbrane}

Next, we consider the localization of the Elko zero mode on Minkowski thick branes.
As we know, there are various thick branes and their properties are also different with each other. We just consider these thick branes embedded in asymptotically AdS space-time. The majority of Minkowski thick brane solutions lead to this case,
such as the solutions with a single scalar field, non-minimally coupled scalar field and so on \cite{DeWolfe,Abdyrakhmanov:2005fs,Gremm:1999pj,Afonso:2006gi,Kehagias:2000au,Bazeia:2008zx,Dzhunushaliev:2009va,Bogdanos:2006qw,Guo:2011wr,Liu:2012gv}. As examples, we just review two solutions: one is for a standard scalar field \cite{Dzhunushaliev:2009va,Gremm:1999pj,Afonso:2006gi} and the other is for a scalar field non-minimally coupled to the Ricci scalar curvature \cite{Dzhunushaliev:2009va,Bogdanos:2006qw,Guo:2011wr,Liu:2012gv}.

The thick brane action of a standard scalar coupled to gravity can be written as
\begin{eqnarray}
S={\int}d^{5}x\sqrt{-g}\left[\frac{1}{2}R-\frac{1}{2}(\partial\phi)^{2}-V(\phi)\right].
\end{eqnarray}
For the sine-Gordon potential
\begin{eqnarray}
V(\phi)=\frac{3}{2}c^{2}[3b^{2}\cos^{2}(b\phi)-4\sin^{2}(b\phi)],
\end{eqnarray}
and the Minkowski brane metric (\ref{the line-element for 5D space time}),
the solution is given by \cite{Dzhunushaliev:2009va,Gremm:1999pj,Afonso:2006gi}
\begin{eqnarray}
 \text{e}^{A(y)}&=&\left[{\cosh(cb^{2}y)}\right]^{-1/3b^{2}},  \\
 \phi(y)&=&\frac{2}{b}\arctan \tanh\big(\frac{3}{2}cb^{2}y\big),
\end{eqnarray}
where $b$ and $c$ are parameters related to the brane thickness.

In addition, Refs. \cite{Dzhunushaliev:2009va,Bogdanos:2006qw,Liu:2012gv} considered thick brane solutions of a scalar field non-minimally coupled to the Ricci scalar curvature, and the action is given by
\begin{eqnarray}
S=\int d^{5}x\sqrt{-g}\left[f(\phi)R-\frac{1}{2}(\partial\phi)^{2}-V(\phi)\right],
\end{eqnarray}
where $f(\phi)$ is a function of the scalar field $\phi$. The above action is conformally related to the Einstein frame action with the Ricci scalar term $\frac{1}{2}R$ via the conformal transformation $g_{MN}\rightarrow 2\widetilde{g}_{MN}f(\phi)$. With the coupling function
\begin{eqnarray}
f(\phi)=\frac{1}{2}(1-\xi\phi^{2})
\end{eqnarray}
and the metric (\ref{the line-element for 5D space time}), for a non-zero coupling constant $\xi\neq0$, the solution is given by \cite{Dzhunushaliev:2009va,Bogdanos:2006qw,Liu:2012gv}
\begin{eqnarray}
 \text{e}^{A(y)}&=&\big[\cosh(ay)\big]^{-\gamma},\\
 \phi(y)&=&\phi_{0}\tanh(ay),
\end{eqnarray}
where the $\gamma=2(\frac{1}{\xi}-6)$, and $\phi_{0}=a^{-1}\phi(0)=\sqrt{\frac{3(1-6\xi)}{\xi(1-2\xi)}}$. The parameter $\xi$ satisfies $0<\xi<1/6$, which means that the $\gamma>0$.

\begin{figure}[htb]
\includegraphics[width=8cm,height=6.5cm]{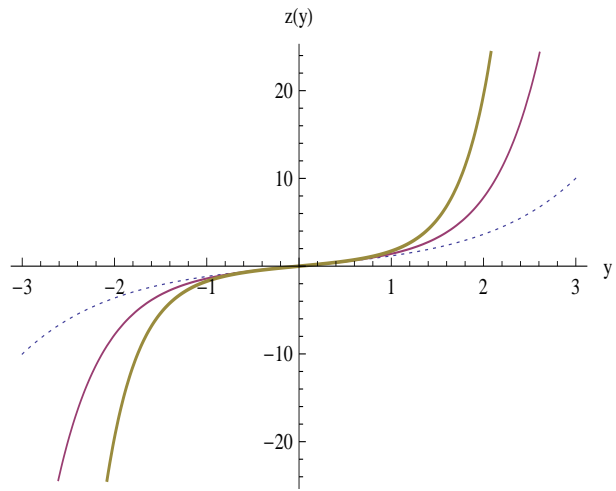}
\caption{ The shapes of the function $z(y)$.
The parameters are set to $a=1$, $b=1$ for dashed line, $b=2$ for thin line, and $b=3$ for thick line.}
\label{Vy}
\end{figure}

We write the warp factors of the two solutions in a unified form
\begin{eqnarray}
\text{e}^{2A(y)}=\cosh(a y)^{-2b}, \label{warp factor 1}
\end{eqnarray}
where $b$ is a positive real constant and $a$ an arbitrary constant parameter. We use (\ref{warp factor 1}) to analyze the localization of the Elko zero mode on these thick branes. The warp factor $\text{e}^{2A(y)}$ is a function of the extra coordinate $y$. But Eq. (\ref{KKequationforzeromode}) should be expressed with the conformally flat coordinate $z$. So, we need the relation between $z$ and $y$, which is related by the coordinate transformation (\ref{coordinate transformation}) and given by
\begin{eqnarray}
z(y)=&-&{i}\frac{\sqrt{\pi}\Gamma(\frac{1+b}{2})}{2|a|\Gamma(1+\frac{b}{2})}\nonumber\\
     &+&{i}\text{sign}(a y)\frac{[\cosh(a y)]^{1+b}}
    {a(1+b)}F,
\label{zy}
\end{eqnarray}
where $F$ is the hypergeometric function
\begin{eqnarray}
F={}_{2}F_{1}\left[\frac{1}{2},\frac{1+b}{2},\frac{3+b}{2},\cosh^{2}(a y)\right].
\end{eqnarray}
Here we face the difficulty that for general $a$ and $b$ we can not get an analytical form of $y(z)$ from the function $z(y)$ given in (\ref{zy}). But we can write the $V_{0}$ (\ref{effective potential Vz}) as a function of $y$:
\begin{eqnarray}
V_{0}(z(y))=\text{e}^{2A}
                  \left(\frac{3}{2}{\partial_y^{2}A}
                  +\frac{19}{4}\left({\partial_y A}\right)^{2}\right).
\end{eqnarray}
As is shown in Fig. \ref{Vy}, we can find that $z(y)$ is a monotonic function. It means that $V_{0}(z)$ has the similar shape and property to $V_{0}(z(y))$.

Now we consider the massless mode $\alpha_0(z)$. Eq. (\ref{KKequationforzeromode}) can be written in the extra coordinate $y$ as
\begin{eqnarray}
\left[-\text{e}^{2A}\partial_{y}^{2}
 -\text{e}^{2A}A'\partial_{y}+V_{0}(z(y))\right]\alpha_0(z(y))=0,
\end{eqnarray}
where the zero mode $\alpha_0(z(y))$ will have the similar figure and property to $\alpha_0(z)$. Let $\alpha_0(z(y))=\text{e}^{-\frac{1}{2}A(y)}\rho(y)$, the above equation is reduced to
\begin{eqnarray}
  \left[-\partial_{y}^{2}
        +5a^{2}b^{2}
        -a^{2}b(2+5b)\text{sech}^{2}(a y)
  \right]\rho(y)=0. \label{Schrodinger equation for zero mode}
\end{eqnarray}
The general solution is given by
\begin{eqnarray}
\rho(y)=C_{1}P_{q-1}^{\sqrt{5}b}(\tanh(a y))+C_{2}Q_{q-1}^{\sqrt{5}b}(\tanh(a y)), \label{solution in case 1}
\end{eqnarray}
where $C_{1}$, $C_{2}$ are integral parameters, $q(q - 1)=b(2+5b)$, $P$ and $Q$ are the first and second Legendre functions, respectively. Hence, we get the solution of the massless mode
\begin{eqnarray}
 \alpha_0(y)=&&\cosh^{b/2}(a y)
      \big[ C_{1}P_{q-1}^{\sqrt{5}b}(\tanh(a y))\nonumber\\
             &&+C_{2}Q_{q-1}^{\sqrt{5}b}(\tanh(a y))
      \big]. \label{chi0 in case 1}
\end{eqnarray}
For arbitrary $b>0$, $\cosh^{b/2}(a y)$ will be divergent when $y\rightarrow\pm\infty$. So the orthonormality condition (\ref{orthonormality relation for zero mode}) requires that $\rho(y)$ should vanish when $y\rightarrow\pm\infty$ if we want get a bound state. From the solution (\ref{solution in case 1}), $\rho(y)$ is a summation of two Legendre functions. According to the theory of the special functions, we know that the Legendre functions $P_{q-1}^{\sqrt{5}b}(\tanh(a y))$ and $Q_{q-1}^{\sqrt{5}b}(\tanh(a y))$ are convergent only under some strong restrictions. For the first Legendre function $P$, it requires that $q-1$ and $\sqrt{5}b$ are integers, or $q-\sqrt{5}b$ or $q-1-\sqrt{5}b$ is zero or negative integer just while $\text{Re}(\sqrt{5}b) < 0$. For the second Legendre function $Q$, it requires that both $q-1$ and $\sqrt{5}b$ are positive half odd integers when $\text{Re}(\sqrt{5}b) > 0$, or $q-1-\sqrt{5}b$ is a negative integer but $\sqrt{5}b$ is not an integer while $\text{Re}(\sqrt{5}b) < 0$. Here we can solve the equation $q(q-1)=b(2+5b)$ and get $q=\frac{1}{2}(1\pm\sqrt{1+8b+20b^{2}})$. Obviously, the solution can not converge at $y=\pm\infty$ since these strong restrictions can not be satisfied. Thus we can not get a bound Elko zero mode. So the zero mode (the 4D massless Elko) of a 5D free massless Elko field can not be localized on these Minkowski thick branes. The result is very interesting. These thick branes which we consider here have the similar asymptotical behavior with the RS$\mathrm{II}$ brane when $z\rightarrow\infty$ and will become RS$\mathrm{II}$ {brane} when the thickness of branes approach $0$. But the localization of the Elko zero mode on these thick {branes} is very different from the one on RS$\mathrm{II}$ brane. If we consider the $V_0$ in Eq. (\ref{KKequationforzeromode}) as a ``potential", it will be a volcano potential for these thick branes. As we know there exist a minimum for a volcano potential, i.e, the depth of the potential well is finite. There will exist a bound zero mode only if the shape of the potentia {is appropriate} (which is depended on the warp factor and the coefficients of the ${A'}^2$ and the $A''$). But for case of the RS$\mathrm{II}$ brane, there exists a delta function at the location of the brane. With the delta function, we have an infinitely deep potential well so that there always exists a bound Elko zero mode. {This is the reason for the difference of the localization of the Elko zero mode on thin and thick branes.}

\subsection{The localization of the massive KK modes of a 5D free massless Elko}

Next we consider the localization of the massive KK modes of 5D free Elko. The mass spectrum of the massive modes is generally expected because it can characterize the geometry of the extra dimension and gives some novel effects coming from the \text{extra dimension}, which may be observed by experiments in the future. It also may give us a new viewpoint to comprehend the origin of the mass of Elko if the massive KK modes can be localized on branes. Thus we should give a detailed analysis of Eq. (\ref{KKequationforElko}) on various kinds of Minkowski branes and we hope to get a reasonable mass spectrum. Before we solve {Eq.} (\ref{KKequationforElko}) in some concrete models, we can speculate that the solution should be a complex function just like a wave function because of the imaginary unit in Eq. (\ref{KKequationforElko}). So it is difficult to get the complete bound massive KK modes.

\subsubsection{The thin brane}
\label{KKmodeonthinbrane}
First, as the typification of thin brane models, we still consider the RS$\mathrm{II}$ model and Eq. (\ref{KKequationforElko}) could read as
\begin{eqnarray}
&&-\alpha_{n}''+\left(\frac{19k^2}{4(1+k|z|)^2}-\frac{3k\delta(z)}{1+k|z|}\right)\alpha_{n}\nonumber\\
&=&\left(m_{n}^2+\text{i}m_{n}\frac{k\text{sign}(z)}{1+k|z|}\right)\alpha_{n}.\label{KKmodeinRSmode}
\end{eqnarray}
And the general solution is
\begin{eqnarray}
\alpha_{n}(z)&=&C_1H(z)M_{\frac{1}{2},-\sqrt{5}}\left(\text{i}\frac{2m_{n}}{k}(k|z|+1)\right)\nonumber\\
             &+&C_2H(-z)M_{\frac{1}{2},\sqrt{5}}\left(\text{i}\frac{2m_{n}}{k}(k|z|+1)\right)\nonumber\\
             &+&C_3W_{\frac{1}{2},-\sqrt{5}}\left(\text{i}\frac{2m_{n}}{k}(k|z|+1)\right),
\end{eqnarray}
where $H(z)$ is the step function and $M$ and $W$ are the two kinds of Whittaker functions, respectively, $C_1$, $C_2$ and $C_3$ are integral parameters and $C_1 M_{\frac{1}{2},-\sqrt{5}}\left(\text{i}\frac{2m_{n}}{k}\right)= C_2 M_{\frac{1}{2},\sqrt{5}}\left(\text{i}\frac{2m_{n}}{k}\right)$ . Let $\alpha_{n}(z)=R_{n}(z)+\text{i}I_{n}(z)$ and the orthonormality condition (\ref{orthonormality relation 1}) requires $\int dz\left(R_{n}^2+I_{n}^2\right)=1$. According to the theory of the special functions, the $M_{\frac{1}{2},\pm\sqrt{5}}\left(\text{i}\frac{2m_{n}}{k}(k|z|+1)\right)$ and $W_{\frac{1}{2},-\sqrt{5}}\left(\text{i}\frac{2m_{n}}{k}(k|z|+1)\right)$ are always divergent. It is obvious that the solutions can not satisfy the orthonormality condition (\ref{orthonormality relation 1}). Thus for any mass $m_{n}$ we can not get a bound massive KK mode of {a} 5D free Elko on the brane in the RS$\mathrm{II}$ model. The result is the same with the one of the scalar field.


\subsubsection{The thick brane}
\label{KKmodeonthickbrane}
We still chose (\ref{warp factor 1}) as the unified form of the  warp factor in thick brane models. For simplicity we just discuss the {case of $b=1$, for which we have $az=\sinh(ay)$, and Eq. (\ref{KKequationforElko}) is read as}
\begin{eqnarray}
&&-\alpha_{n}''+\frac{a^2\left(-6+19(az)^2\right)}{4\left(1+(az)^2\right)^2}\alpha_{n}\nonumber\\
&=&\left(m_{n}^2+\text{i}m_{n}\frac{a^2z}{1+(az)^2}\right)\alpha_{n}.\label{KKmodeinthickmode}
\end{eqnarray}
It is difficult to get a general analytical solution. But we can consider the asymptotic behaviors of the above equation. When $z\rightarrow\infty$ the $A'(z)$ and $A''(z)$ will have the similar behaviors with the ones in the condition of RS$\mathrm{II}$ model. Thus Eq. (\ref{KKmodeinthickmode}) has the similar asymptotic behaviors with (\ref{KKmodeinRSmode}) and the solution should have the similar property. Thus the result should be the same with the case of RSII brane, and for this kind of thick branes, the massive Elko KK modes can not be localized on the branes. This analysis is reasonable because these thick branes {are all} embedded in asymptotically AdS space-time.

{ We can also achieve the same conclusion from the following analysis. The massive KK mode $\alpha_n(z)$ should be a complex function according to Eq. (\ref{KKequationforElko}). Let $\alpha_n(z)=R_n(z)+\text{i}I_n(z)$ with $R_n(z)$ and $I_n(z)$ the real functions, then Eq. (\ref{KKequationforElko}) is reduced to
\begin{eqnarray}
&&\!~~R_n''\!-\!\Big(\frac{3}{2}A''\!+\!\frac{13}{4}(A')^2\!-\!m_n^2\Big)R_n\!+\!m_n A'I_n\nonumber\\
&+&\!\text{i}\Big[I_n''-\!\Big(\frac{3}{2}A''\!+\!\frac{13}{4}(A')^2\!-\!m_n^2\Big)I_n\!-\!m_n A'R_n\Big]=0.~~~
\end{eqnarray}
Thus we get the following coupled equations:
\begin{eqnarray}
-R_n''+V_{e}R_n-m_n A'I_n&=&m_n^2R_n,\label{KKequationforR}\\
-I_n''~+V_{e}I_n+m_n A'R_n&=&m_n^2I_n.\label{KKequationforI}
\end{eqnarray}
Here $V_{e}=\frac{3}{2}A''+\frac{13}{4}(A')^2$. In this paper, we just consider the RSII brane and the thick branes embedded in asymptotically AdS space-time, for which we have the $A'\rightarrow0$ when $z\rightarrow\infty$. Thus when $z\rightarrow\infty$, the terms $m_n A'I_n$ and $m_n A'R_n$ in Eqs. (\ref{KKequationforR}) and (\ref{KKequationforI}) will vanish so that Eqs. (\ref{KKequationforR}) and (\ref{KKequationforI}) can be approximated as
\begin{eqnarray}
-R_n''+V_{e}R_n&=&m_n^2R_n,\\
-I_n''~+V_{e}I_n~&=&m_n^2I_n.
\end{eqnarray}
They are the Schr\"{o}dinger-like equations and the effective potential $V_{e}$ is a volcano potential with vanishing value at the boundary of the extra dimension for these branes embedded in AdS or asymptotically AdS space-time. As we know, for such a volcano potential, there do not exist bound massive KK modes. Thus all the massive KK modes can not be normalized and hence can not be localized on these branes. This is some what like the cases of a 5D free massless scalar field and Dirac spinor field. Physically, we can understand this as the following explain. Usually, a brane has a power of confining matter fields. Concretely, the localization is decided by some factors such as the structure of the brane, the number of the dimensions of extra dimensions, the mass of the KK modes of the matter fields, the coupling way between the matter fields and the background field generating the brane. We have known that usually the zero mode of a 5D free massless scalar field can be localized on the Minkowski brane embedded in 5D AdS or asymptotically AdS space-time, but the zero modes of a 5D free massless vector and a 5D Dirac spinor can not. For these branes, the massive modes of the free scalar, vector and fermion can not be localized. However, if we consider de-Sitter or Anti-de-Sitter branes, or introducing the coupling terms between the matter fields and the background field, the matter fields may be localized on the branes. Since the Elko field has some characteristics of both scalar and Dirac spinor, it is not strange that its massive KK modes can not be localized on these Minkowski branes considered in this paper. Here, even there is an imaginary unit in the dynamical equation of the KK modes of the Elko field, the result is also similar to the cases of the scalar and Dirac spinor.}

\section{Localization of 5D Elko spinors with coupling term on Minkowski branes}
\label{section4}
In this section, we will study the localization of Elko spinors with coupling term on Minkowski branes in 5D space-time. As is shown in last section, for a free Elko field in 5D space-time, we can not get any bound massive KK mode on Minkowski branes and even can not get a bound Elko zero mode on Minkowski thick branes. As the localization of Dirac spinors on branes, we introduce here the interaction between the Elko spinor and the background scalar, and the simplest choice is the Yukawa coupling. The corresponding Lagrangian density is
\begin{eqnarray}
\mathfrak{L}_{\text{Elko}}=&-&\frac{1}{4}g^{MN}\left(\mathfrak{D}_{M}\mathop \lambda\limits^\neg\mathfrak{D}_{N}\lambda+\mathfrak{D}_{N}\mathop \lambda\limits^\neg\mathfrak{D}_{M}\lambda\right) \nonumber\\
&-&\eta F(\phi)\mathop \lambda\limits^\neg\lambda, \label{Lagrangian for Elko with V in 5D}
\end{eqnarray}
where the $F(\phi)$ is a function of the background scalar field $\phi$ and $\eta$ is the coupling constant.
When $F(\phi)$ is a constant, the additional term in (\ref{Lagrangian for Elko with V in 5D}) is a mass term with  $M_{Elko}^2=\eta F(\phi)$.
Then the equation of motion for the Elko field coupled with the scalar is read as
\begin{eqnarray}
\frac{1}{\sqrt{-g}}\mathfrak{D}_{M}(\sqrt{-g}g^{MN}\mathfrak{D}_{N}\lambda)-2\eta F(\phi)\lambda=0. \label{Elko's motion equation 2}
\end{eqnarray}
By considering the conformally flat metric (\ref{conformally flat line-element}), using the non-vanishing components of the spin connection for the flat branes (\ref{spin connection for flat brane}), introducing the KK decomposition (\ref{Elkodecomposition}), and noticing {the} linear independance of {the}  $\varsigma^{(n)}_{+}$ and $\tau^{(n)}_{+}$ ($\varsigma^{(n)}_{-}$ and $\tau^{(n)}_{-}$), we can obtain the equation for the KK {mode} $\alpha_{n}$:
\begin{eqnarray}
&&-\alpha_{n}''+\left(\frac{3}{2}A''+\frac{13}{4}(A')^{2}+2\eta\text{e}^{2A}F(\phi)\right)\alpha_{n}\nonumber\\
&=&\left(m_{n}^2-\text{i}m_{n}A'\right)\alpha_{n}. \label{KKequationforElkowithV}
\end{eqnarray}
When we just pay our attention to the zero mode of the 5D Elko, $m_0=0$ and Eq. (\ref{KKequationforElkowithV}) can be read as
\begin{eqnarray}
[-\partial_{z}^{2}+V_{0}(z)]\alpha_{0}(z)=0, \label{Schrodinger equation 2}
\end{eqnarray}
where
\begin{eqnarray}
V_{0}(z)=\frac{3}{2}A''+\frac{13}{4}{A'}^{2}+2\eta \text{e}^{2A}F(\phi). \label{effective potential Vz 2}
\end{eqnarray}
As discussed in the last section, if we want to get the action of the 4D massless and massive Elko spinors from the action of a 5D Elko with coupling term:
\begin{eqnarray}
S_{\text{Elko}}&=&\int d^5x\sqrt{-g}\big[-\frac{1}{4}g^{MN}\left(\mathfrak{D}_{M}\mathop \lambda\limits^\neg\mathfrak{D}_{N}\lambda+\mathfrak{D}_{N}\mathop \lambda\limits^\neg\mathfrak{D}_{M}\lambda\right) \nonumber\\
                &&-\eta F(\phi)\mathop \lambda\limits^\neg\lambda\big]\nonumber\\
               &=&-\frac{1}{2}\sum_{n}\int d^4x \left(\partial^{\mu}\hat{{\mathop \lambda\limits^\neg}}^{n}\partial_{\mu}\hat{\lambda}^{n}+m_{n}^2\hat{{\mathop \lambda\limits^\neg}}^{n}\hat{\lambda}^{n}\right),
\end{eqnarray}
we should introduce the orthonormality condition (\ref{orthonormality relation 1}).

\subsection{The localization of the zero mode of 5D Elko with coupling term}
\label{secVolcanoPotentialswithV}

First we still consider the localization of the zero mode of a 5D Elko spinor with coupling term. As we have emphasized, 5D free {massless} Elko fields can not be localized on Minkowski thick branes. As we know there exist many similarities between the Elko field and the scalar field. Here, let us consider the equation of motion of a 5D {free} massless scalar field. It turns out for Minkowski branes to be
\begin{eqnarray}
\partial^{\mu}\partial_{\mu}\Phi+\text{e}^{-3A}\partial_{z}(\text{e}^{3A}\partial_{z}\Phi)=0,
 \label{motion equation of scalar field}
\end{eqnarray}
from which one can investigate 4D scalar fields by the KK decomposition $\Phi=\sum_n \phi_n(x)h_n(z)\text{e}^{-3A/2}$ and obtain the Schr\"{o}dinger-like equation for the scalar KK modes $h_n$ \cite{Liu:2009dwa,Liu:2008wd}:
\begin{eqnarray}
[-\partial_{z}^{2}+V_{\Phi}]h_n=m_{n}^{2}h_n,
\end{eqnarray}
where the effective potential $V_{\Phi}$ is given by
\begin{eqnarray}
V_{\Phi}(z)=\frac{3}{2}A''+\frac{9}{4}{A'}^{2}. \label{PotentialVzScalarKKmodes}
\end{eqnarray}
The zero mode of a 5D {free} massless scalar can be localized because the corresponding Schr\"{o}dinger-like equation can be factorized, which is the result of the fact that the coefficient of ${A'}^{2}$ is the square of the coefficient of $A''$. Obviously, for Minkowski thick branes, the difference of the coefficients of the ${A'}^2$ between the $V_{0}$ and $V_{\Phi}$ would prevent the factorization of the $V_{0}$ and hence prevent the localization of the Elko zero mode. When an appropriate $F(\phi)$ is introduced, the coefficient of the ${A'}^2$ may be adjusted to be the same as the one of the scalar case, so that there exists the bound Elko zero mode. Thus we assume
\begin{eqnarray}
V_{0}(z)&=&\frac{3}{2}A''+\frac{13}{4}{A'}^{2}+2\eta \text{e}^{2A}F(\phi)\nonumber \\
              &=&\frac{3}{2}A''+\frac{9}{4}{A'}^{2}, \label{PotentialVzElkoKKmodeswithYokawa}
\end{eqnarray}
which is identical to
\begin{eqnarray}
 \big(\partial_{z}A(z)\big)^{2}+2\eta \text{e}^{2A(z)}F(\phi)=0.
\end{eqnarray}
It is more clear when the {above} equation is written in extra coordinate $y$:
\begin{eqnarray}
 \big(\partial_{y}A(y)\big)^{2}=-2\eta F(\phi(y)). \label{lationshipFandA}
\end{eqnarray}
This equation depends on the warp factor $\text{e}^{2A(y)}$, the scalar field $\phi$ and the function $F(\phi)$. It is reasonable to consider the scalar-Elko coupling $\eta\mathop \lambda\limits^\neg\phi^n\lambda$. Hence $F(\phi)$ can be taken as $\phi^n$ and $F(\phi(y))$ should be an even function of $y$ according to Eq. (\ref{lationshipFandA}). As we know, for majority of the brane models, the scalar field $\phi$ is a kink, i.e., it is an odd function of $y$, so the simplest case is $n=2$. Then we have:
\begin{eqnarray}
 \big(\partial_{y}A(y)\big)^{2}=-2\eta\phi^2(y), \label{lationshipphiandA1}
\end{eqnarray}
or
\begin{eqnarray}
 \partial_{y}A(y) \propto\phi(y). \label{lationshipphiandA2}
\end{eqnarray}
If the warp factor $\text{e}^{2A(y)}$ and the scalar field $\phi$ are related by Eq. (\ref{lationshipphiandA2}), the Elko zero mode, i.e., the 4D massless Elko particle may be localized on the brane. It is exciting that there are many models  satisfying this relation (\ref{lationshipphiandA2}). We will discuss them in the following subsections respectively.

\subsubsection{The thin brane}
\label{thinbrane2}

First, we still consider the RS$\mathrm{II}$ model. From Eq. (\ref{actionofRS}), it is clear that there does not exist a background scalar field $\phi$ in this model. Here we introduce the 5D mass term, i,e, $\eta\phi^2=M_{Elko}^2$ with
 $M_{Elko}$ the 5D Elko mass.
At the same time, notice that $A'(y)=-k$ $\text{sign}(y)$ and ${A'}^2(y)=k^2$. It is natural that
\begin{eqnarray}
M_{Elko}^2\propto k^2. \label{zeromodeforRS}
\end{eqnarray}
For an arbitrary constant $M_{Elko}$, the $V_{0}$ (\ref{effective potential Vz 2}) is read as:
\begin{eqnarray}
V_{0}&=&\frac{19k^2}{4(1+k|z|)^2}
  +\frac{2M_{Elko}^2}{(1+k|z|)^2}-\frac{3k\delta(z)}{1+k|z|}\nonumber \\
  &=&\frac{(19+8\epsilon)k^2}{4(1+k|z|)^2}-\frac{3k\delta(z)}{1+k|z|}.
\end{eqnarray}
Here $\epsilon=M_{Elko}^2/k^2$. We can get the bound Elko zero mode by solving the equation (\ref{Schrodinger equation 2}):
\begin{eqnarray}
\alpha_{0}(z)=\sqrt{(-1+\sqrt{5+2\epsilon})k}~(1+k|z|)^{\frac{1}{2}-\sqrt{5+2\epsilon}}.
\end{eqnarray}
Here it is required that $\epsilon>-2$. So, for any $M_{Elko}^2\ge0$, the Elko zero mode can be localized on the RS$\mathrm{II}$ brane.


\subsubsection{The thick brane}
\label{Tnickbrane2}

Generally, for thick brane models based on the {General Relativity}, Eq. (\ref{lationshipphiandA2}) can not be satisfied. Fortunately it can be satisfied in some thick branes within modified gravity theory, for example, the thick brane solution of a scalar field non-minimally coupled to the Ricci scalar curvature \cite{Dzhunushaliev:2009va,Bogdanos:2006qw,Liu:2012gv}. The solution is $\text{e}^{A(y)}=(\cosh(ay))^{-\gamma}$ and $\phi(y)=\phi_{0}\tanh(ay)$. Obviously, the derivative of the warp factor $A(y)$ and the scalar field are related by
\begin{eqnarray}
A'(y)&=& -\gamma\partial_{y}\ln[\cosh(a y)]\nonumber \\
     &=& -a\gamma\tanh(a y)=-\frac{a\gamma}{\phi_{0}}\phi(y).
\end{eqnarray}
Let $\eta=-\frac{a^2\gamma^2}{2\phi_{0}^2}$ then Eq. (\ref{lationshipphiandA1}) is satisfied.

\begin{figure}[htb]
\includegraphics[width=8cm,height=6.5cm]{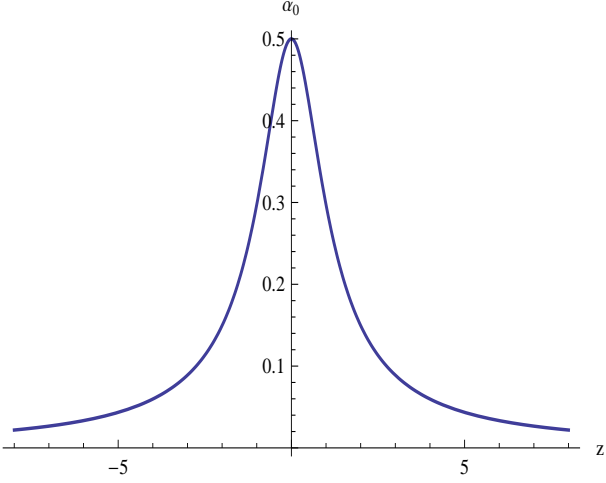}
\caption{ The shape of
the Elko zero mode $\alpha_{0}(z)$ (\ref{Elkozeromodeonthickbrane})
in the thick brane model with a scalar field non-minimally coupled to the Ricci scalar curvature. The parameters are set to $\gamma=1$ and $a=1$.}
\label{VandX}
\end{figure}

When Eq. (\ref{lationshipphiandA1}) is satisfied, the $V_{0}$ (\ref{effective potential Vz 2}) can be rewritten as
(\ref{PotentialVzElkoKKmodeswithYokawa}). After factorizing Eq. (\ref{Schrodinger equation 2}), we have
\begin{eqnarray}
\left[\partial_{z}+\frac{3}{2}A'(z)\right]\left[-\partial_{z}+\frac{3}{2}A'(z)\right]\alpha_{0}(z)=0.
\end{eqnarray}
Then we get the bound Elko zero mode:
\begin{eqnarray}
\alpha_{0}(z)=C\text{e}^{\frac{3}{2}A(z)},
\end{eqnarray}
where $C$ is the normalization constant.
For the thick brane model \cite{Dzhunushaliev:2009va,Bogdanos:2006qw,Liu:2012gv}, the zero mode is
\begin{eqnarray}
\alpha_{0}(z(y))=C\cosh(a y)^{-\frac{3}{2}\gamma}.
\end{eqnarray}
The normalization constant $C$ depends on the parameters $\gamma$ and $a$. When $\gamma=1$, $az=\sinh(ay)$ and the normalized zero mode reads
\begin{eqnarray}
\alpha_{0}(z)=\frac{a}{2}[1+(a z)^{2}]^{-\frac{3}{4}}. \label{Elkozeromodeonthickbrane}
\end{eqnarray}
The shape of the above Elko zero mode is plotted in Fig. \ref{VandX}.
It is clear that for the thick brane models satisfying Eq. (\ref{lationshipphiandA2}), the Elko zero mode can be localized on the thick branes if ($\mathrm{I}$) the coupling term $\eta \phi^2 \mathop \lambda\limits^\neg\lambda$ is introduced and the coupling constant $\eta$ is taken as some particular expression determined by the parameters in the models, and ($\mathrm{II}$) the Elko zero mode is normalizable.

\subsection{The localization of the massive KK modes of a 5D Elko with coupling term}

Next we consider the localization of the massive KK modes of a 5D Elko with coupling term. As we have shown in the last section, the massive KK modes of a 5D free Elko can not be localized on Minkowski branes. We expect that the massive KK modes of a 5D Elko can be localized on branes by introducing a coupling term as in the case of a Dirac field. We just discuss the simple and meaningful coupling term $\eta\mathop \lambda\limits^\neg\phi^n\lambda$ which we have considered in the case of the zero mode. It is foreseeable that the coupling term may not change the result about the localization of the massive KK modes because the equation of the KK modes (\ref{KKequationforElkowithV}) will have the similar asymptotic behaviors to the one of a 5D free Elko (\ref{KKequationforElko}). We still investigate the localization of the massive KK modes in different kinds of brane models respectively.

\subsubsection{The thin brane}
\label{thinbrane3}

In subsection {\ref{thinbrane3}}, we introduce the 5D mass term $M_{Elko}^2\mathop \lambda\limits^\neg\lambda$ to the action of a 5D Elko in RS$\mathrm{II}$ model. For an arbitrary constant $M_{Elko}$, Eq. (\ref{KKequationforElkowithV}) is given by
\begin{eqnarray}
&&-\alpha_{n}''+\left(\frac{(19+8\epsilon)k^2}{4(1+k|z|)^2}-\frac{3k\delta(z)}{1+k|z|}\right)\alpha_{n}\nonumber\\
&=&\left(m_{n}^2+\text{i}m_{n}\frac{k\text{sign}(z)}{1+k|z|}\right)\alpha_{n}, \label{KKequationforElkowithVinRS}
\end{eqnarray}
where $\epsilon=M_{Elko}^2/k^2$. It is obvious that the asymptotic behavior is the same with (\ref{KKmodeinRSmode}) and the general solution is
\begin{eqnarray}
\alpha_{n}(z)&=&C_1H(z)M_{\frac{1}{2},-\sqrt{5+2\epsilon}}\left(\text{i}\frac{2m_{n}}{k}(k|z|+1)\right)\nonumber\\
             &+&C_2H(-z)M_{\frac{1}{2},\sqrt{5+2\epsilon}}\left(\text{i}\frac{2m_{n}}{k}(k|z|+1)\right)\nonumber\\
             &+&C_3W_{\frac{1}{2},-\sqrt{5+2\epsilon}}\left(\text{i}\frac{2m_{n}}{k}(k|z|+1)\right),
\end{eqnarray}
For any $\epsilon${,} the Whittaker functions $M_{\frac{1}{2},\pm\sqrt{5+2\epsilon}}\left(\text{i}\frac{2m_{n}}{k}(k|z|+1)\right)$ and $W_{\frac{1}{2},-\sqrt{5+2\epsilon}}\left(\text{i}\frac{2m_{n}}{k}(k|z|+1)\right)$ are always divergent. Thus we still can not get bound massive KK modes of a 5D Elko on the brane by introducing a 5D mass term in the RS$\mathrm{II}$ model.

\subsubsection{The thick brane}
\label{thickbrane3}

For simplicity we just investigate the localization of the massive KK modes of a 5D Elko with coupling term on the thick brane of a scalar field non-minimally coupled to the Ricci scalar curvature \cite{Dzhunushaliev:2009va,Bogdanos:2006qw,Liu:2012gv}. We introduce the scalar-Elko coupling $\eta\mathop \lambda\limits^\neg\phi^n\lambda$ and let $\gamma=1$, {for which we have $az=\sinh(ay)$.} { Thus the equation (\ref{KKequationforElkowithV}) is reduced to
\begin{eqnarray}
&&-\alpha_{n}''+\left(\frac{a^2\left(-6+19(az)^2\right)}{4\left(1+(az)^2\right)^2}+\frac{2\eta(az)^n}{\left(1+(az)^2\right)^{\frac{n}{2}+1}}\right)\alpha_{n}\nonumber\\
&=&\left(m_{n}^2+\text{i}m_{n}\frac{a^2z}{1+(az)^2}\right)\alpha_{n}.\label{KKmodeinthickmodewithV}
\end{eqnarray}
When $z\rightarrow\infty$, the additional term ${2\eta(az)^n}/{\left(1+(az)^2\right)^{\frac{n}{2}+1}}$, which comes from the coupling term, has the same asymptotic behavior with ${2\eta}/{(az)^2}$ and so it can be combined with the original term ${a^2\left(-6+19(az)^2\right)}/{4\left(1+(az)^2\right)^2}$. Hence the solution should be similar with (\ref{KKmodeinthickmode}) and the result will not be changed. Thus there does not exist any bound Elko massive KK mode on this thick brane by introducing the scalar-Elko coupling $\eta\mathop \lambda\limits^\neg\phi^n\lambda$.} In fact it is foreseeable that, for any thick brane embedded in asymptotically AdS space-time and the kink scalar field $\phi$, the $\eta\phi^n$ will be a constant when $z\rightarrow\infty$ so the additional term $2\eta\text{e}^{2A}F(\phi)=2\eta\text{e}^{2A}\phi^n$ in {Eq.} (\ref{KKequationforElkowithV}) will vanish and the asymptotic behavior will be the same with (\ref{KKequationforElko}). The {situation} may change when we introduce {a} special coupling term for example $F(\phi)=\frac{1}{(\phi_0^2-\phi^2)^k}$. When we consider the thick brane solution of a scalar field non-minimally coupled to the Ricci scalar curvature \cite{Dzhunushaliev:2009va,Bogdanos:2006qw,Liu:2012gv} and let the $\gamma=1$, $k=2$, {Eq.} (\ref{KKequationforElkowithV}) can be read as
\begin{eqnarray}
&&-\alpha_{n}''+\left(\frac{a^2\left(-6+19(az)^2\right)}{4\left(1+(az)^2\right)^2}+2\eta(1+(az)^2)\right)\alpha_{n}\nonumber\\
&=&\left(m_{n}^2+\text{i}m_{n}\frac{a^2z}{1+(az)^2}\right)\alpha_{n}.\label{KKequationforElkowithV2}
\end{eqnarray}
When $z\rightarrow\infty$ {Eq.} (\ref{KKequationforElkowithV2}) {has} the same behavior with
\begin{eqnarray}
-\alpha_{n}''+2\eta(1+(az)^2)\alpha_{n}=m_n^2\alpha_n.
\end{eqnarray}
This is just a Schr\"{o}dinger-like equation with an infinite potential well and we can get a series of bound KK modes and a mass spectrum. But it does not make sense to introduce the coupling term $\frac{\eta}{(\phi_0^2-\phi^2)^k}\mathop \lambda\limits^\neg\lambda$ (the coupling term $\eta\mathop \lambda\limits^\neg\phi^n\lambda$ can be explained as the Elko spinors coupled to $n$ scalar particles, but the $\frac{\eta}{(\phi_0^2-\phi^2)^k}\mathop \lambda\limits^\neg\lambda$ is unreasonable). Thus we can not get any bound massive KK mode of 5D Elko by introducing a meaningful coupling term.

\section{The candidate of dark matter}
\label{section5}

Elko is regarded as a natural dark matter candidate for its interesting properties \cite{Ahluwalia:2004ab,Ahluwalia:2004sz}. First, a 4D Elko is a spin-1/2 fermionic field with mass dimension one{,} which is very different from  mass dimension 3/2 associated with the conventional Dirac fermionic field. Obviously the mismatch of the mass dimensions prevents Elko to enter the fermionic doublets of {SM} and it can be used to explain the dark matter interacting {very} weakly with SM matters and electromagnetic radiations.

Second, the dark matter is self-interacting, which is suggested by observational evidences. As we know, for the usual Dirac fermionic field, the self-interaction will be suppressed by the Planck scale. For the scalar field which {carries} mass dimension one the suppression will not happen so the scalar field is also a candidate of dark matter. The {case} is the same to Elko. Thus in 4D space-time, the self-interaction of dark matter can be described by  the following self-interaction of Elko \cite{Ahluwalia:2010zn}{:}
\begin{eqnarray}
g_{\Lambda}[{\mathop \Lambda\limits^\neg}(x)\Lambda(x)]^{2},
\end{eqnarray}
where $g_{\Lambda}$ is {a} dimensionless coupling {constant} and $\Lambda(x)$ is the quantum field of the Elko {spinors} (in fact it has two types). {On the other hand, when we choose a fermionic dark matter, there will be {an} important advantage.  A fermionic dark matter {makes} it possible to support the dark matter halo by Fermi degenerate pressure.} In Refs. \cite{Ahluwalia:2004ab,Ahluwalia:2010zn}, the authors even gave the following relationship between the Elko mass and Chandrasekhar value for the halo's size:
\begin{eqnarray}
R_{ch}\sim{x}_{Elko}^{-2}6.3\times10^{-2}pc,
\end{eqnarray}
where $x_{Elko}$ is the Elko mass $m$ in unit of keV. From the relationship we can infer $m \sim 1$eV.

For 4D Elko {spinors}, the only coupling with the SM {fields} which will not be unsuppressed is {the one} with the Higgs:
\begin{eqnarray}
g_{\phi\Lambda}\phi^{\dagger}(x)\phi(x){\mathop \Lambda\limits^\neg}(x)\Lambda(x),
\end{eqnarray}
where $g_{\phi\Lambda}$ is a dimensionless coupling constant and $\phi(x)$ the SM Higgs doublet \cite{Ahluwalia:2010zn}. As we have emphasized, for {a} 5D Elko, only the zero mode (massless 4D Elko spinor) can be localized successfully on Minkowski branes. The conclusion hints that the coupling with Higgs which can generate the mass of 4D Elko {spinors} is crucial. And there {exists} a chance to obverse the interaction between Elko and Higgs in LHC, which was investigated in Ref. \cite{Dias:2010aa}.

Third, for the dark matter, it is proved that the dark matter couples to an axis,
which has come to be known as the axis of evil. As a non-local field,
Elko field will be a local quantum field along a preferred axis,
which is in the direction perpendicular to the Elko plane \cite{Ahluwalia:2008xi,Ahluwalia:2009rh}, and was proposed to the axis of locality in the dark sector.
This is also an advantage of choosing Elko as the candidate of dark matter.
Thus Ahluwalia and Grumiller suggested that Elko can be considered as a first-principle candidate of dark matter.
All these advantages motive us to investigate the localization of Elko on branes.


\section{Conclusion and discussion}
\label{section6}

In this paper, we have investigated the localization of a 5D Elko field on Minkowski branes. First, we briefly reviewed some fundamental structures of Elko, which show that Elko is very different from a usual Dirac fermionic field but it is similar to a scalar field on many aspects--the Lagrangian density and mass dimension, for example.
Then, we considered various kinds of Minkowski branes and analyze the localization of {a} 5D Elko field on these branes by presenting the equation of {the} Elko KK modes.

The 5D Elko field was investigated in two cases: the 5D massless free Elko field and the 5D Elko field with coupling term. In the first case, we found that the Elko zero mode can be localized on the RS$\mathrm{II}$ brane but can not be localized on the majority of Minkowski thick branes {embedded in asymptotically AdS space-time}. There {does} not exist any bound massive KK mode on these Minkowski branes.

In the second case, the Elko zero mode still can be localized on the RS$\mathrm{II}$ brane for any 5D massive Elko, and it can also be localized on some specific thick branes if certain coupling term such as ${\eta} F(\phi){\mathop\lambda\limits^\neg}\lambda$ is introduced. {However}, we {still} can not find any bound massive KK mode on these branes by introducing a meaningful coupling term.

For these thick branes generated by a kink scalar, in order to localize the 5D Dirac field, we usually introduce the coupling term $\eta \phi^{2n-1}\overline{\psi}\psi$ ($n$ is a positive integer), and when the coupling constant $\eta>\eta_{0}$ ($\eta_{0}$ is some constant parameter decided by the branes), then the left or right chiral Dirac fermion zero mode can be localized on these branes \cite{Liu:2007ku,Zhang:2007ii,Zhao:2009ja,Li:2010dy,Castro:2010uj,Chumbes:2010xg,Zhao:2010mk,Liu:2010pj,Brihaye:2008am,Liu:2009dw,Liu:2009uca,Ringeval:2001cq,Koley:2004at,Davies:2007tq,Liu:2009mga,Liu:2009ve}. But for {a} 5D Elko field, in order to localize the zero mode, the similar coupling term is changed as $\eta \phi^{2n}{\mathop\lambda\limits^\neg}\lambda$ and the coupling constant $\eta$ should be taken as some particular expression determined by the parameters in the models. No matter what the exponent of the $\phi$ is chosen as, we can not find any bound massive KK mode of {a} 5D Elko on these branes. Thus there may only exist the bound zero mode of a 5D Elko on Minkowski branes and the conclusion hints that the coupling with Higgs which can generate the mass of 4D Elko spinors is crucial.

In addition, for {a} Dirac field, only {the} left or right chiral fermion zero mode can be localized on these branes. But for Elko, it is interesting to find that the KK modes of different types of Elko are the same and only the zero modes can be localized on the branes.

There are still some issues. First, we just considered Minkowski {branes} {embedded in AdS space-time} and can not find any bound massive KK mode of a 5D Elko. Maybe the conclusion will be changed when we consider some other Minkowski thick branes, for example, Weyl thick Branes \cite{Arias:2002ew,BarbosaCendejas:2005kn,BarbosaCendejas:2006hj}. Second, in order to investigate the localization of 4D massive Elko fields, some new mechanisms need to be introduced. It is difficult to find  bound massive KK modes because of the imaginary unit in Eq. (\ref{KKequationforElko}). But we may find some Elko resonances like the case of Dirac field. The resonances for Elko should be redefined because Eq. (\ref{KKequationforElko}) is not a Schr\"{o}dinger-like equation. Third, it is interesting to investigate the localization of a 5D Elko on de Sitter and anti-de Sitter branes. We leave these problems in the further works.

%
%

\acknowledgments
The authors would like to thank the anonymous referees whose comments largely helped us in improving the original manuscript, and thank Professors Dharam Ahluwalia, Sebastian Horvath, Hao Wei as well as doctors Xiao-Long Du, Zhen-Hua Zhao, Zhi-Feng Sun and Shao-Wen Wei for helpful discussions.
This work was supported in part by the Huo Ying-Dong Education Foundation of
Chinese Ministry of Education (No. 121106), the National Natural
Science Foundation of China (No. 11075065), the Doctoral Program Foundation of Institutions of Higher Education of China
(No. 20090211110028), and the Fundamental Research Funds for the
Central Universities (No. lzujbky-2012-k30).

\end{document}